\documentclass{egpubl}
 
\JournalSubmission    
\usepackage[T1]{fontenc}
\usepackage{dfadobe}  

\usepackage{cite}  
\usepackage[labelfont=bf,textfont=it]{caption}
\usepackage{comment}
\usepackage{amsmath}
\usepackage{amssymb} 
\usepackage{amsfonts}
\usepackage{multirow}
\usepackage{color}
\usepackage{ifthen}
\usepackage{xcolor}
\usepackage[normalem]{ulem} 
\newcommand{\etal}{{\textit{et~al.}}}
\usepackage{enumitem}
\usepackage[encapsulated]{CJK} 
\usepackage{wrapfig}
\usepackage{soul}
\usepackage{subcaption}

\definecolor{gray}{rgb}{0.5,0.5,0.5}
\definecolor{green}{rgb}{0, 0.6, 0}
\definecolor{orange}{rgb}{1, 0.5, 0}
\definecolor{mahogany}{rgb}{0.75, 0.25, 0.0}
\definecolor{purple}{rgb}{0.6, 0, 0.6}
\definecolor{darkgreen}{rgb}{0, 0.3, 0}
\definecolor{orange}{rgb}{1, 0.5, 0.}

\newboolean{revising} 
\setboolean{revising}{true}
\ifthenelse{\boolean{revising}}
{
\newcommand{\ignore}[1]{}
\newcommand{\nothing}[1]{}

\newcommand{\note}[1]{\textcolor{red}{#1}}
\newcommand{\myhl}[1]{#1}
\newcommand{\cgfhl}[1]{#1}

\newcommand{\iccmt}[1]{\textcolor{purple}{ichao: #1}}

}
{
\newcommand{\note}[1]{}
\newcommand{\iccmt}[1]{}

\newcommand{\ignore}[1]{}
\newcommand{\nothing}[1]{}
}

\usepackage{algorithm}
\usepackage{algorithmicx}
\usepackage{algpseudocode}
\algnewcommand\algorithmicinput{\textbf{Input:}}
\algnewcommand\INPUT{\item[\algorithmicinput]}
\algnewcommand\algorithmicoutput{\textbf{Output:}}
\algnewcommand\OUTPUT{\item[\algorithmicoutput]}
\algnewcommand\algorithmicforeach{\textbf{for each}}

\algdef{S}[FOR]{ForEach}[1]{\algorithmicforeach\ #1\ \algorithmicdo}
\algrenewcommand{\alglinenumber}[1]{\color{red!80!blue}\footnotesize#1:}

\algnewcommand\Func[2]{\textcolor{green}{#1}\textcolor{green}{(#2)}}
\algnewcommand\Insert[2]{Insert {#1} to #2.}
\algnewcommand\Input[1]{\State \textbf{Input: } #1}
\algnewcommand\Output[1]{\State \textbf{Output: } #1}

\newcommand{\ie}{i.e.,}
\newcommand{\eg}{e.g.,}
\newcommand{\figname}{Figure}

\newcommand{\secname}{Section}

\newcommand{\eqname}{Eq.}

\BibtexOrBiblatex
\electronicVersion
\PrintedOrElectronic
\ifpdf \usepackage[pdftex]{graphicx} \pdfcompresslevel=9
\else \usepackage[dvips]{graphicx} \fi

\usepackage{egweblnk} 

\title[Interactive Optimization of Generative Image Modeling using Sequential Subspace Search and Content-based Guidance]
      {Interactive Optimization of Generative Image Modeling using Sequential Subspace Search and Content-based Guidance} 

\makeatletter
\renewcommand*{\@fnsymbol}[1]{\ifcase#1\or*\else\@arabic{\numexpr#1-1\relax}\fi}
\makeatother

\author[Toby Chong et al.]
{\parbox{\textwidth}{\centering Toby Chong Long Hin$^{1*}$ ~ I-Chao Shen$^{1,2}$\orcid{0000-0003-4201-3793}\thanks{Both authors contributed equally to the paper.}~
        Issei Sato$^{1}$ ~ Takeo Igarashi$^{1}$
        }
        \\
{\parbox{\textwidth}{\centering $^{1}$The University of Tokyo ~ $^{2}$National Taiwan University
       }
}
}

\begin{document}
\teaser{
 \includegraphics[width=\linewidth]{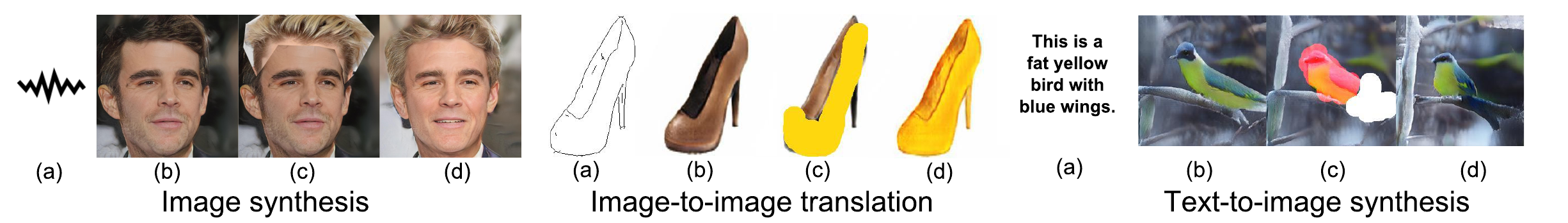}
 \centering
  \caption{    
  Our system provides additional control to a pretrained and frozen generative image modeling network without modifying the network itself. 
    It first generates an image (b) from input signal (a). The user edits the image, and the result (c) is then used as guidance for optimization. The user obtains the final result (d) after a few iterations of slider control.}
\label{fig:teaser}
}

\maketitle
\begin{abstract}
Generative image modeling techniques such as GAN demonstrate highly convincing image generation result. 
However, user interaction is often necessary to obtain desired results. Existing attempts add interactivity but require either tailored architectures or extra data.
We present a human-in-the-optimization method that allows users to directly explore and search the latent vector space of generative image modeling.
Our system provides multiple candidates by sampling the latent vector space, and the user selects the best blending weights within the subspace using multiple sliders.
In addition, the user can express their intention through image editing tools. 
The system samples latent vectors based on inputs and presents new candidates to the user iteratively.
An advantage of our formulation is that one can apply our method to arbitrary pre-trained model without developing specialized architecture or data.
We demonstrate our method with various generative image modeling applications, and show superior performance in a comparative user study with prior art iGAN~\cite{iGAN2016}.


\end{abstract}
\section{Introduction}
\label{sec:intro}
With the growing maturity of 
generative image modeling techniques such as generative adversarial networks~\cite{goodfellow2014generative} (GANs) and auto-encoder architecture~\cite{VAE}, many applications have been proposed, including high quality face generation~\cite{Karras2018ICLR}, image inpainting~\cite{IizukaSIGGRAPH2017}, image generation from text~\cite{attngan}, and image style transfer and translation~\cite{pix2pix,CycleGAN,MUNIT}.
These applications produce images from a random latent vector and additional inputs (\eg~a sketch or a sentence) that exhibit characteristics observed in the training data.

A common problem with these approaches is the lack of controllability of the synthesized results.
When the user detects a defect in a synthesized result (\eg~artifacts, unwanted image attributes), it is often difficult to fix that problem.
Some existing approaches allow interactive control of the image synthesis process, \eg~using color point cues for image colorization~\cite{zhangColorization}, or 
sketches as structural cues for image translation~\cite{pix2pix} and face image editing~\cite{faceshop}.
However, several challenges remain.
First, the network architecture is usually tailored to allow only specific controls.
Second, the tailored network must be trained using appropriate training data.
\myhl{iGAN~\mbox{\cite{iGAN2016}} addressed these challenges by building a natural image manifold and allowing the users to explore the manifold by drawing. 
Although their approach is straightforward, we observed that novice users often failed to render the desired images accurately.
Hence, using raw stroke input is not sufficient for generating the desired images.}

In this paper, we address these challenges by developing a generic framework imparting controllability to a pretrained model.
We consider existing networks and models as a fixed black box function; it is not necessary to introduce novel network architectures or use additional training data to provide additional controllability. 
This makes our method applicable to a board range of existing works as well as future architectures, which is critical as image generative model is evolving in an immense pace.
We treat the problem of improving generative image modeling as an optimization problem in the input latent vector space.
The user guides optimization via slider control and image editing tools.
Slider control allows the user to explore the subspace around the desired point in the latent vector space.
Image editing tools allow the user to directly indicate desired changes to the system, such as to the colors used, and to identify regions that should be preserved and problematic regions that should be removed.

Our method is based on human-in-the-loop optimization, which was originally introduced for parameter tweaking during image-editing~\cite{Koyama:2017:SLS}.
However, it is difficult to efficiently obtain the desired result using the original sequential line search method, because our target space is much larger (around 512d) than the earlier space (less than 10d).
Therefore, we propose two extensions to the existing method. 
First, we use multiple sliders rather than a single slider to allow the user to explore a larger subspace than the 1-D subspace of the original method.
This allows the system to attain an optimum more efficiently. 
Second, we introduce a content-aware sampling strategy that favors the results indicated by user editing. 
We achieve this by formulating different image operations as content-aware bias term, and add it to the acquisition function which is a function used in Bayesian optimization to sample the next candidates.

We demonstrate the effectiveness and versatility of our proposed framework using three different image synthesis applications: face synthesis, image translation, and text-to-image synthesis, together with other user studies and ablation studies.

To summarize, the key contributions of this paper are:
\begin{itemize}
\item A generic framework imparting controllability to generative image modeling without requiring tailoring of the network architecture or training.
\item Sequential subspace search effectively exploring the high-dimensional latent vector space.
\item Content-aware sampling strategy that favors the synthesized results suggested by the user.
\item User studies comparing our method with iGAN~\cite{iGAN2016}. 
These results demonstrate the effectiveness of the slider-based exploration over a pure drawing interface.
\end{itemize}




\section{Related Work}
\label{sec:related}
\subsection{Interactive Generative Image Modeling}
Many researchers have developed independent methods incorporating different user inputs to generate outputs created by generative image modeling.
One promising approach is the use of conditional GAN to condition different inputs (such as labels or aerial images) to solve the image-image translation problem~\cite{pix2pix} and sketching of terrain authoring system~\cite{Guerin:2017:IET}.
Zhang~\etal~\shortcite{zhangColorization} created additional local and global hint networks to incorporate local and global user inputs.
Portenier~\etal~developed FaceShop~\shortcite{faceshop}, a novel network architecture combining both image completion and translation in a single framework.
The user draws strokes, using both geometry and color constraints to guide face editing.
However, all these methods require tailored network architectures and training data to endow specific applications with controllability. 
We address this problem by developing a generic framework introducing additional controllability to a pretrained model.

\cgfhl{
There is another line of research that shares similar goal with us, \ie~to provide controllabilities over image generative modeling by exploring the latent space constructed by pre-trained deep generative models.
\mbox{Zhu~\etal~\shortcite{zhu2016generative}} proposed iGAN that enables the users to create image content by drawing line sketches and paint colors on blank canvas.
\mbox{Bau~\etal~\shortcite{GANPaint}} proposed an interactive system for semantic image editing tasks with limited semantic labels such as removing windows, add tress and so on.
These semantic labels are discovered \mbox{GANDissect~\cite{bau2019gandissect}} and requires additional analysis for each generative image modeling applications.
\mbox{Gu~\etal~\shortcite{gu2020image}} proposed a new inversion method that operate on multiple latent vectors instead of one.
Their novel inversion method enable multiple image processing applications such as image colorization, inpainting and so on.
\mbox{Abdal~\etal~\shortcite{abdal2019image2stylegan, abdal2020image2stylegan++}} proposed an image editing framework based on an  image inversion method tailored for \mbox{StyleGAN~\cite{karras2019style}}.
}

\cgfhl{
A clear difference between the aforementioned methods and ours is that ours is designed to ease user inputs by introducing an additional and simpler model of interaction (\ie~slider). 
This is critical in situations where the user intention cannot be clearly expressed (\ie~the user drawing do not portrait the intent).
Although the recent proposed iterative inversion methods obtain better reconstruction results compared to iGAN, we argue that compares to iGAN is more suitable since iGAN combines iterative optimization inversion with user controls.
The other methods requires additional training for each input image~\mbox{\cite{GANPaint}}, decision of the number of latent codes~\mbox{\cite{gu2020image}} or tailored latent space for exploration~\mbox{\cite{abdal2020image2stylegan++}}.
Nevertheless, these advances on using different loss functions or optimizers can benefit the interactive methods including our method and iGAN.
}


\cgfhl{
To validate the efficacy of our method, we conducted two user studies to compare the results of our method and iGAN both perceptually (using real image) and numerically (using image generated by GAN).
And we show that our method is able to generate image closer to reference image both perceptually and numerically.
}

\subsection{Bayesian Optimization with Gaussian Process}\label{related_bo}
Bayesian optimization (BO) is a framework designed to optimize expensive-to-evaluate black-box functions with minimal number of evaluations:
\begin{align}
    \max_{x} f(x),
\end{align}
where $f(x)$ is a black-box function with unknown derivatives and convexity properties.
During each trial, BO chooses a sample point based on previous observations; more specifically, BO optimizes an acquisition function using a predefined prior; in this work we focused on BO with a Gaussian Process (GP) as the prior, which is commonly used for such task~\cite{Koyama:2017:SLS}.
The acquisition function seeks the next sample point that meet a criterion such as expected improvement (EI)~\cite{EI}, knowledge gradient (KG)~\cite{KG}, or variations thereof; please refer to ~\cite{shahriari2016taking} for details.
As BO effectively approximates arbitrary functions, it is widely used to explore high-dimensional parameter spaces, such as hyperparameter tuning systems for machine learning algorithms~\cite{Hyperopt, BAYE_HYPER_DNN}, photo enhancement~\cite{Koyama:2017:SLS}, material BRDF design~\cite{Brochu:2007b, Koyama:2017:SLS}.
\mbox{REMBO~\cite{rembo-wang2016bayesian}} proposed a random embedding to reduce dimensions and
has been proven effective on high dimensional (1 billion) problem with low effective dimension.

Bayesian optimization with inequality constraints has been proposed~\cite{Gardner:ICML2014,Gelbart:UAI2014}, for scenarios where the feasibility cannot be determined in advance.
In this problem setting, they incorporate
inequality constraints into a black-box optimization:
\begin{align}
    \max_{x} f(x)~\mathrm{s.t.}~ c(x)\leq \epsilon,
\end{align}
where $f(x)$ and $c(x)$ are the results of some expensive experiments, i.e., black-box functions.
We formulated the inequality constraints by incorporating a feasibility indicator function into the Bayesian optimization process.


\section{Overview}
\label{sec:overview}

\begin{figure}[t!]
\centering
\includegraphics[width=\linewidth]{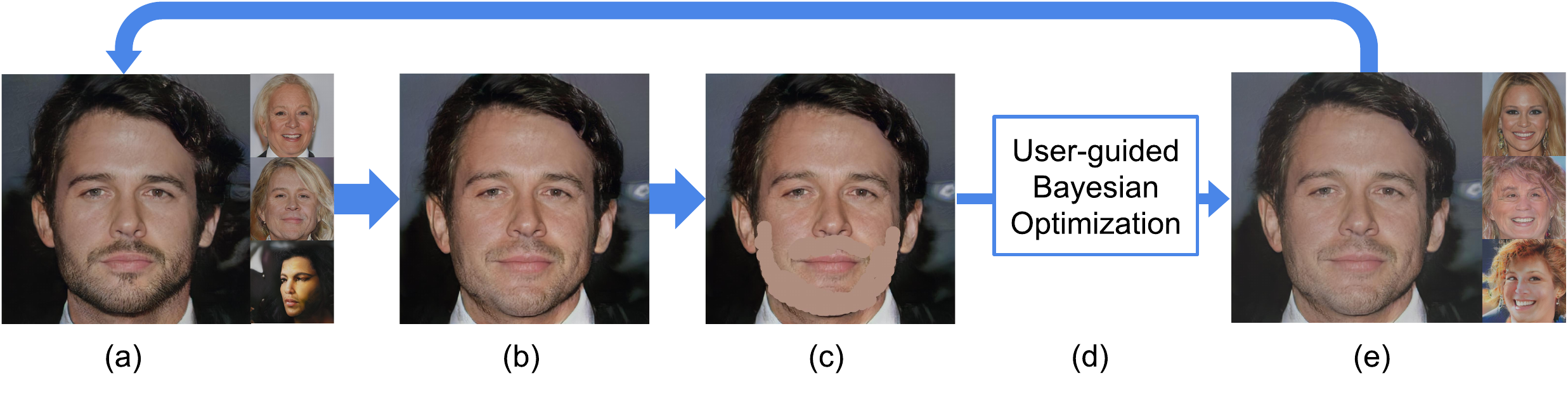}
\caption{
\textbf{Illustration of our workflow to remove beard.}
At each iteration, the user starts with initial images (a) (supplied by the user or randomly generated). 
They manipulate the sliders (b) and edits on the blended image (c).
Our user-guided Bayesian Optimization (d) 
generates three new candidates (e) for the user to explore and generate new blending image. 
}
\label{fig:workflow}
\end{figure}

\figname~\ref{fig:workflow} shows the workflow of our framework.
The system first shows an image $I^1_0$ (\figname~\ref{fig:workflow}(a)) generated with a random vector $z^1_0$ using frozen pretrained generator network $G$, \ie~$I^1_0 = G(z^1_0)$, and $I_j^i$ refers to the $i$-th candidate in the $j$-th iteration.
However, the user may not be satisfied for its defects or personal preference on $I^1_0$; such as hair blending in with background and unshaven beard respectively.
Assume that the user prefers a face similar to the current image $I^1_0$, but without the defect and the beard.
The user first adjusts the multi-way slider to obtain an image ${I}'_0=G({z}'_0)$ (\figname~\ref{fig:workflow}(b)) without defects by blending generated candidate images $\{I_0^i=G(z^i_0)\}_{i=1}^{c}$, where $c$ is the number of candidates ($c=4$ in our current implementation).
Furthermore, our framework provides several image-editing tools as in iGAN allowing the user to engage in additional
guidance (\secname~\ref{sec:ui}).

The editing tools allow the user to directly edit ${I}'_0$ by painting over it, or sourcing external images guidance image ${I}^{*}_0$ (\figname~\ref{fig:workflow}(c)). 

The user presses the \textit{next} button and the system provides the next set of candidate images. 
We use ${z}'_0$ as $z_1^1$ and the new latent vectors $\{z^i_1\}_{i=2}^{c}$ to generate the candidate images $\{I^i_1\}_{i=1}^{c}$ for the next user input iteration.
This iterative process continues until the user obtains the image that matches his/her preferences.

\section{User interface}
\label{sec:ui}
\begin{figure}
\centering
\includegraphics[width=0.8\linewidth]{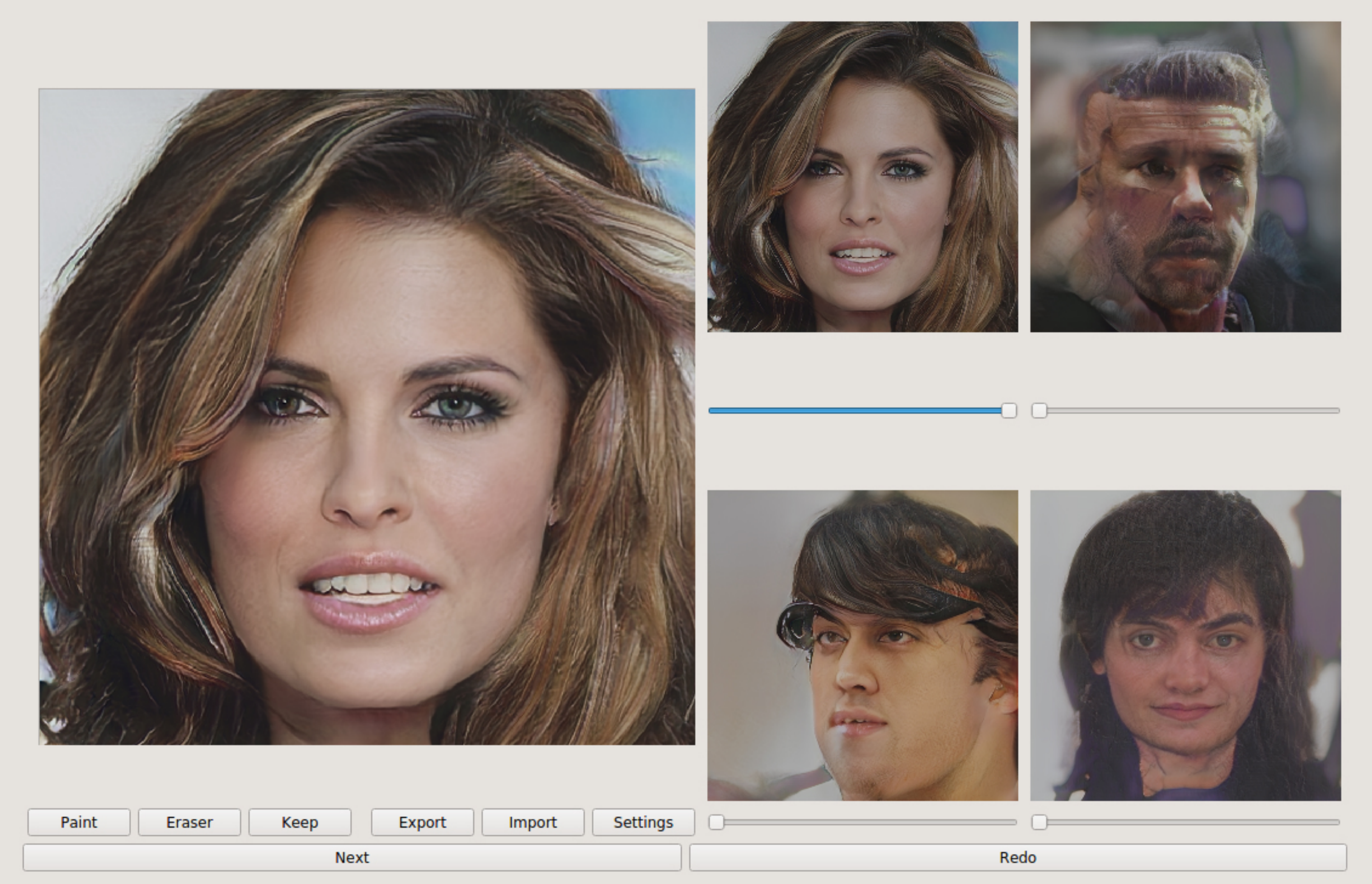}
\caption{
Screenshot of our user interface. 
It allows users to control the synthesis process by controlling slider and image-editing operations, including \textit{painting} to achieve desired colors, \textit{eraser} erase undesired regions, \textit{keep} preserve desire regions, and \textit{export} and \textit{import} to \textit{copy \& paste} using an external editor.
After providing guidance, the user presses the \textit{next} button to obtain the next candidates.
}
\label{fig:ui_screenshot}
\end{figure}

Our user interface consists of a main viewing window and multiple candidate images for user selection as shown in \figname~\ref{fig:ui_screenshot}.
For each candidate image, we provide an associated user slider.
Adjusting the slider enables the user to explore and compose a new image $\tilde{I}$ that is shown in the principal window.
Moreover, we provide several local image editing tools to allow for further user guidance.
At each iteration, the user can specify preferences both by manipulating the sliders and performing local edits.
For example, the user can manipulate the second slider to obtain a friendly face, and paint the skin white to brighten it.
Finally, the user hits the ``next'' button to request the system to update the internal model and present the next candidates.

\subsection{Multi-way slider}
At iteration $t$, the user manipulates the sliders to compose a new image (${I}'_t$) that is close to his/her preferences.
The slider values correspond to blending weights of candidate images; the user explores the convex subspace of the latent vector space bounded by the candidate images.
If the user is not satisfied with the blended image ${I}'_t$, he/she can use the image editing tools to obtain a final guidance image $I^*_t$ (\secname~\ref{sec:user_tool}).
Otherwise, he/she can directly use the blended image as a guidance image, \ie~$I^*_t={I}'_t$, to obtain new candidate images. 

\subsection{Image editing tools}
\label{sec:user_tool}
Image editing tools allow the user to deliver requests directly to the system. We currently provide the following editing tools.
\begin{description}[style=unboxed,leftmargin=0cm]
\item[Color painting]
This tool assigns desired colors to the blended image (${I}'_t$).
The user paints arbitrary regions; our algorithm finds candidates that match the assigned colors inside the painted region.
This tool is similar to those of existing systems~\cite{adobephotoshop18}.
\item[Eraser]
If the user is not satisfied with the content of an ROI within the blended image (${I}'_t$), but can not explain why, he/she can simply erase the ROI.
Our system next provides candidates that look as different as possible in terms of the erased region.
\item[Keep]
This tool allows the user to specify a region on $I'$ to be preserved in the next iteration. 
This is the opposite of the eraser.
\item[Copy \& Paste]
This tool allows the user to employ an existing image editor of choice to adjust the image.
The user can employ images from any source, \eg~from a web search or image databases, that match his/her preferences for a specific region.
The user then copies and pastes that region from the found image to the current image (${I}'_t$).
The system then tries to find candidates that match the assigned image inside the region.
\end{description}


\section{Method}
\label{sec:method}
Our method focuses on adjusting the input latent vector $z$ in order to control image synthesis process $\mathcal{G}(z)$ of generative image modeling.
We consider this latent vector adjusting process to be a numerical optimization, and model the user preference as our objective function.
Since evaluating such user preference function is expensive-and-hard, we design our method based on Bayesian optimization that aims for optimizing it with minimal number of observations.
In the following paragraphs, we will give a brief introduction about Bayesian optimization and how to do it on subspace search oracle, which enable simple adjustment for human.

\subsection{Sequential Subspace Search}
\label{sec:sss}
Given an initial latent vector $z_0$, which generates an image $I_0 = \mathcal{G}(z_0)$. We seek to maximize $g(I)$, a \textbf{user preference function} describing how much a user prefers an image $I$. 
By nature, $g(I)$ is expensive to evaluate and often inconsistent. 
Therefore, a requirement of our method would be to maximize such function with as little observation as possible. 
Bayesian Optimization (BO) is traditionally deployed for optimizing expensive to evaluate function, see more discussion on BO in \secname~\ref{related_bo}.
Koyama~\etal~\shortcite{Koyama:2017:SLS} extends BO and proposes to use slider interface which allows the users to consistently express their own preferences to a large number of options.
Internally the system uses Bradley-Terry-Luce model~\cite{BTL} to determine the relative numerical preference values of the user, and BO to provide candidate used to form the next slider. 
In this work, we propose multiple sliders user interface, which greatly enhance the flexibility and efficiency to the user compared to the original single slider.
Below we describe the mechanism we used to model user preference function $g(I)$ using multiple sliders.

\begin{description}[style=unboxed,leftmargin=0cm]
\item[User interaction and its mathematical formulation]
\label{user interaction}
Our user interface consists of $c$ sliders, each represents a latent vector $z^{1...c}$ and a corresponding image $I^{1...c}$, a user explore the entire subspace by steering through it with the provided sliders.
Given slider values $\{s^i\}^{i=1}_c$, blending weights $a^i$ for each latent vector $z^i$ are given as $\{a^i={s^i}/{\sum_{j=1}^c s^j}\}_{i=1}^c$. 
Note that we use the normalized slider values, \ie~slider values $(1,1,1,1)$ is equivalent to $(0.5,0.5,0.5,0.5)$. 
The blended image, $I^b = \mathcal{G}(z^b)$, generated by the blended latent vector $z^b = \sum_{j=1}^c a^jz^j$, is updated and displayed on the left side of the user interface as the user manipulate the sliders. 
After the user finish manipulating the sliders, we assume that they arrive to the global minimum of the user preference function within the subspace, \ie~ they choose the best image within the subspace. 
Below we use the phrase ``user preference at latent space'' to refer to the underlying perceptional response $g(I)$ of user observing an image $I = G(z)$ generated by a fixed generator $G$. 
To avoid complicated notation, we use $g(z)$ as a short hand for $g(G(z))$ since we assume a fixed $G$. 
And we refer to $g$ as the goodness values and $g(z)$ as the function itself. The latent space refers to a search space which is defined during the training of $G$.
\item[Modelling user preference]
In order to infer the goodness value $g(z)$ from slider interaction, we formulate this as a maximum a posterior (MAP) estimation. Following sequential line search, we formulate the slider manipulation task with BTL model~\cite{BTL} and user preference at latent space as a Gaussian process prior as we target general generative model inference and possess no domain specific knowledge. Concretely, the objective function that we want to maximise using (MAP) estimation is the following:
\begin{align}
\begin{split}
(g^{\mathrm{MAP}}, \theta^{\mathrm{MAP}}) &= \operatorname*{argmin}_{g,\theta} {p(g,\theta | \mathcal{P}, z^b)}\\
                        &= \operatorname*{argmin}_{g,\theta} p(\mathcal{P}, z^b|g,\theta) p(g|\theta) p(\theta),
\end{split}
\end{align}
where $\theta$ is the parameters of the Gaussian Process (detail in \mbox{Sec~\ref{user_preference_modelling}}), $\mathcal{P}$ is the latent vectors sampled with user interaction, $g^{MAP}$ and $\theta^{MAP}$ are the maximum likelihood estimate of $g$ and $\theta$ respectively.
This optimization serves two purposes, one is to extract numerical value of user preference from slider manipulation (\textit{slider manipulation modelling}).
And another one is to estimate user preference for any unobserved latent variable $z$ (\textit{user preference modelling}).
\end{description}
\subsubsection{Slider manipulation modelling}Given a set of $c$ multidimensional latent variables, collected at iteration $t$, $\mathcal{P}_t = \{z^i_t\}_{i=1}^c$, and the variable corresponding to $z^b_t$ that is chosen by the user, we describe this situation as 
\begin{align}
z^b_t \succ{} \mathcal{P}_t \backslash {z^b_t} 
\end{align}
and it's likelihood using BTL model as 
\begin{align}
p(z^b_t \succ{\mathcal{P}_t} \backslash {z^b_t} | \{g(z^i_t)\}_{i=1}^c) = \frac{exp(g(z^b_t) / s)}{\sum_{i=1}^c exp(g(z^i_t) / s)},
\end{align}
where $s$ is a hyperparameter to adjust the sensitivity of the model.
We set $s=1$ through all our experiments.
\subsubsection{User preference modelling}\label{user_preference_modelling}
We also model the underlying user preference toward the latent space as a Gaussian Process $\mathcal{GP}$. We assume $\mathcal{GP}$ to follow a multivariate Gaussian distribution, parameterized with $\theta$, describing the kernel used in the function. In all of our experiments, we follow Koyama~\etal~\shortcite{Koyama:2017:SLS} and use RBF kernel.
By using such formulation we avoid overly constrain user with the specific domain / application, and also allow inference of user preference of unobserved latent variables.

\cgfhl{As $\mathcal{D}$ and $\theta$ are conditionally independent given $g$, at iteration $t$, we have }
\begin{align}
p(\mathcal{D}, z^b|g,\theta) = p(\mathcal{D}, z^b|g) = \prod_{j=0}^t p({z^b_j} \succ{\{z^i_j\}_{i=1}^c } | g),
\end{align}
where $\mathcal{D}$ is the slider manipulation responses.
We also define the following prior.
\begin{align}
    p(g|\theta) &= \mathcal{N}(g;0,K), \\
    p(\theta_i) &= \mathcal{LN}(\ln{0.5}, 0.1), \\
    p(\theta) &= \prod_i {p(\theta_i)},
\end{align}
where $p(g|\theta)$ is defined to be the $\mathcal{GP}$ prior, and $K$ and $\theta$ are the covariance matrix and the parameters of the kernel (in our experiment we use RBF kernel) and the latent variables $\{\mathcal{P}, z^b\}$. We also construct $g'$, an Gaussian Process regressor that approximates $g$ using $K$ and $\theta$.

\begin{figure}[!t]
\includegraphics[width=\linewidth]{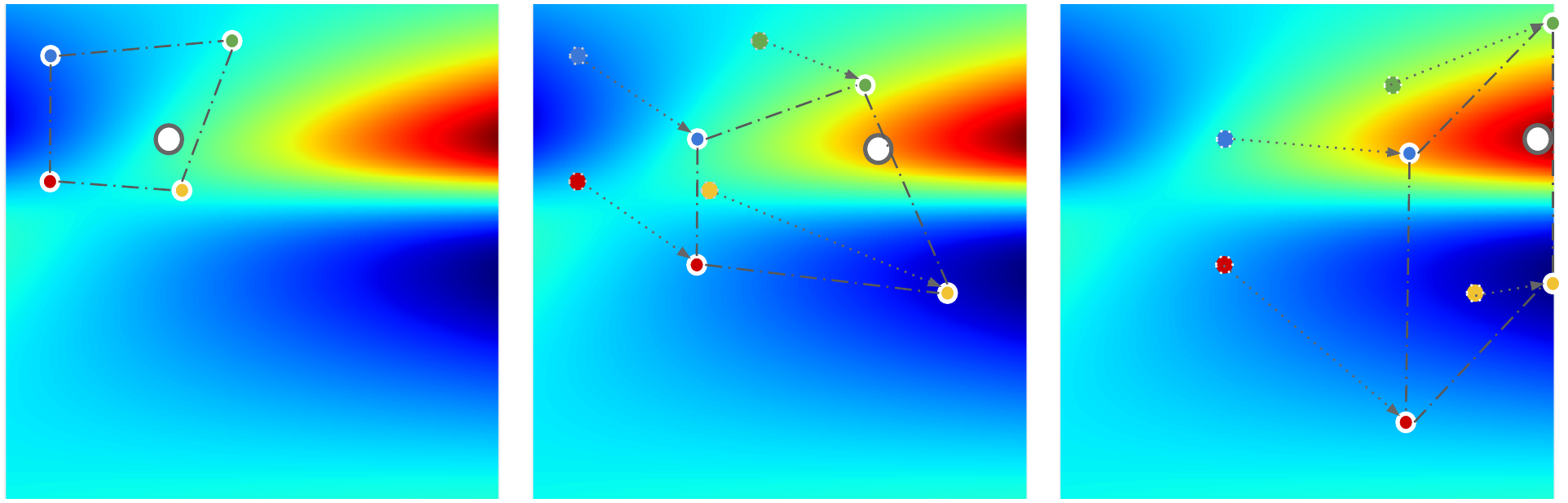}
\caption{
An illustration of the sequential subspace search process using 
a two-dimensional test function.
The iteration proceeds from left to right. 
At each iteration, the user first locates the local maximum (white dot) within the search region, the system then locates the next 3 most probable candidates (colored dots) and form a subspace for the next iteration. 
}
\label{fig:bo_seq}
\end{figure}

For further explanation and implementation detail of Gaussian Process with sequential line search, we advice the reader to read the sequential line search paper~\cite{Koyama:2017:SLS}.
\figname~\ref{fig:bo_seq} illustrates an example optimization sequence of our subspace search with multi-way sliders. 

\subsection{Preference learning by Bayesian optimization}
\label{sec:bo_line}

For iteration $t>1$, we use the chosen latent variable in the last iteration to be the starting latent variable, such that $
z^0_t$ = $z^{\mathrm{chosen}}_{t-1}$. Therefore at iteration $t>1$, we have $m = t \cdot c+1$ samples.
Suppose that we currently have a set of $m$ observations $\mathcal{O}$ for $m$ iterations, containing the latent variables and their estimated user preference values:
\begin{align}
\mathcal{O}_m = \{\mathcal{P}_m, g'(\mathcal{P}_m)\}.
\end{align}
The next observation points $\{z^i_{m+1}\}^c_{i=2}$ should be ``the ones most worth observing'' point based on the all previous observed data $\mathcal{O} = \{\mathcal{O}_n \}^m_{n=1}$.
We define an \textit{acquisition function} $a_{\mathcal{O}}(z)$ to measure the ``worthiness'' of the next sampling candidate $z_{m+1}$.
For each iteration, the system maximizes the acquisition function to determine the next sampling point:
\begin{align}
z_{m+1} = \operatorname*{argmax}_{z\in\mathcal{Z}} {a_{\mathcal{O}}(z)}.
\label{eq:next_sample}
\end{align}
In order to choose the next sampling point which is most worthy to sample from, the \textit{expected improvement} (EI) criterion is often used.
Let $g'^{+}$ be the maximum value among the observation $\mathcal{O}$, the acquisition function is defined as
\begin{align}
a_{\mathcal{O}}^{\mathrm{EI}}(z) = E[\max\{g'(z) - g'^{+}, 0\}],
\label{eq:acq}
\end{align}
where $E[X]$ means the expectation value of $X$. 

\begin{description}[style=unboxed,leftmargin=0cm]
\item[Selection of multiple candidates]
We combine the expected improvement with constant liar strategy \cite{CL} for acquiring multiple points simultaneously.  
We first obtain the first candidate through maximizing the current acquisition function.
Then, we assign the maximum score to this sample point and update the acquisition (e.g. we assume the new candidate is as good as the best candidate we have seen so far).
And we pick the second candidate that maximizes the update acquisition function.
We repeat this process to obtain a set of $n$ candidates.
\end{description}


\begin{figure*}[h!]
    \centering
    \includegraphics[width=\linewidth]{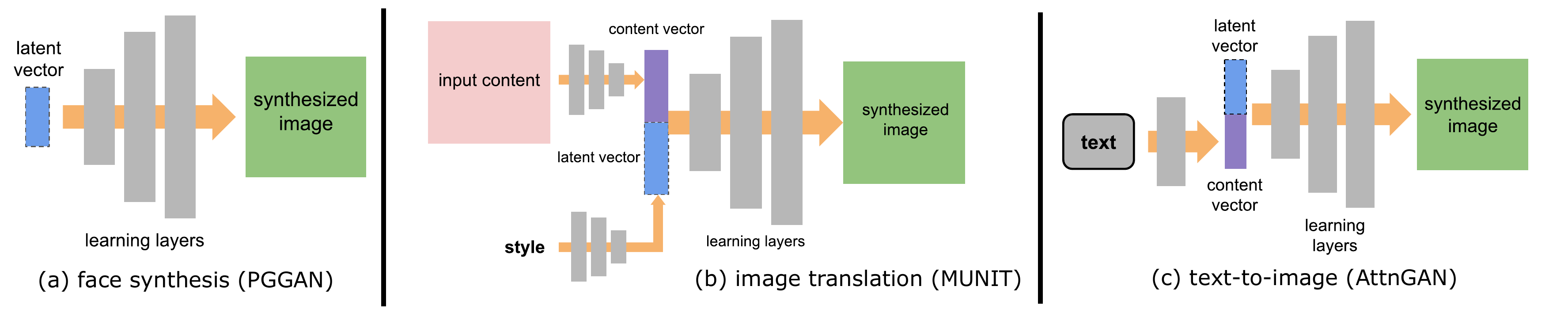}
    \caption{
      Overview of the generator architectures we used to demonstrate our method. 
      The blue part is the latent space in which our framework operates.
    }
    \label{fig:network_arch}
\end{figure*}

\subsection{Content-aware sampling strategy}
\label{sec:guide_acq}
To incorporate the user guidances during the sampling process as described at section ~\ref{sec:user_tool}, we extend the original acquisition function (\eqname~\ref{eq:acq}) into the following form:
\begin{align}
a_{\mathcal{O}}^{\mathrm{cEI}}(z) = E[\max\{g'(z) - g'^+, 0\}] - \sigma_1 C(\mathcal{G}(z)) - \sigma_2 \mathcal{R}(z), 
\label{eq:c_acq}
\end{align}
where $C(I)$ is the content-aware bias term, $\mathcal{R}(z)$ is a regularization term, and $\sigma_1$ and $\sigma_2$ are the weights to balance the scale between the expected improvement, the content-aware bias term, and the regularization term.

Maximizing $a^{\mathrm{cEI}}$ can be regarded as the following constraint optimization problem:
\begin{align}
\max_{z} E[\max\{g'(z)-g'^+,0\}]~\mathrm{s.t.}~C(G(z))\leq \epsilon.    
\end{align}
Note that deciding $\epsilon$ corresponds to adjusting $\sigma$.
The existing Bayesian optimization with inequality constraints deals with a hard constraint and thus multiply expected improvement by a feasibility indicator function.
In this work, we consider a soft constraint and thus add a proximal regularization/constraint as described in the next. 

\begin{description}[style=unboxed,leftmargin=0cm]
\item[Content-aware term]
We handle all the guidance from the image editing operation with the following term:
\begin{align}
C(I) = \sum_i{ \sum_j{ \sum_c{ (I^{*}_{i,j,c} - I_{i,j,c})^2 * M_{i,j,c}}}},
\end{align}

where $M$ is a mask initialized as all $0.2$, as we observed in the pilot study that users mostly prefer to preserve the image content outside ROI.
We use $i,j$ as the 2D pixel coordinate of the image and $c$ as the channel.
For color painting and copy \& paste, we set $M=1$ inside ROI.
It encourages the optimized image to match the user's edit inside ROI.
For the keep operation, we also set $M=1$ inside ROI, which encourages the part of the image marked as keep stay the same in the optimized image. 
For eraser, we set $M=0$ inside ROI to encourage the optimization to alter in the optimized image within the erased region.

It is important to notice that evaluating the content-aware term $C$ requires the evaluation of $G(x)$ which is rather expensive. To this end, we run $n$ L-BFGS optimizations in parallel and combine all the evaluations into a batch to speed up the optimization, we use the best candidate across all the optimizations to be the final result.

\item[Regularization term]
We introduce an regularization term to incoporate our prior knowledge of $z$, \ie~
the latent variables are usually sampled from a known distribution $p$, \eg~Normal distribution.
Therefore we design a regularization term $R$ to prevent the estimation from deviating too far from $p$: $R(z) = log(p(z))$.

\end{description}

\section{Results}
\label{sec:results}
We used our framework to control the synthesis processes of three different applications.
\figname~\ref{fig:network_arch} shows an overview of the network architecture.
We used either pretrained generator networks or networks trained as described in the original papers.
During user interactions, the generator network parameters were frozen.

The users usually require only 1-3 iterations, mostly manipulate the sliders to generate the desired images across all the applications shown in the paper.
This suggests that our content-aware guidance from image editing operations is helpful for finding good candidates for users to explore. 
Please refer to the supplemental materials 2 for complete editing sequences for all the results shown in the paper. 

\begin{figure*}[h!]
\centering
\includegraphics[width=\linewidth]{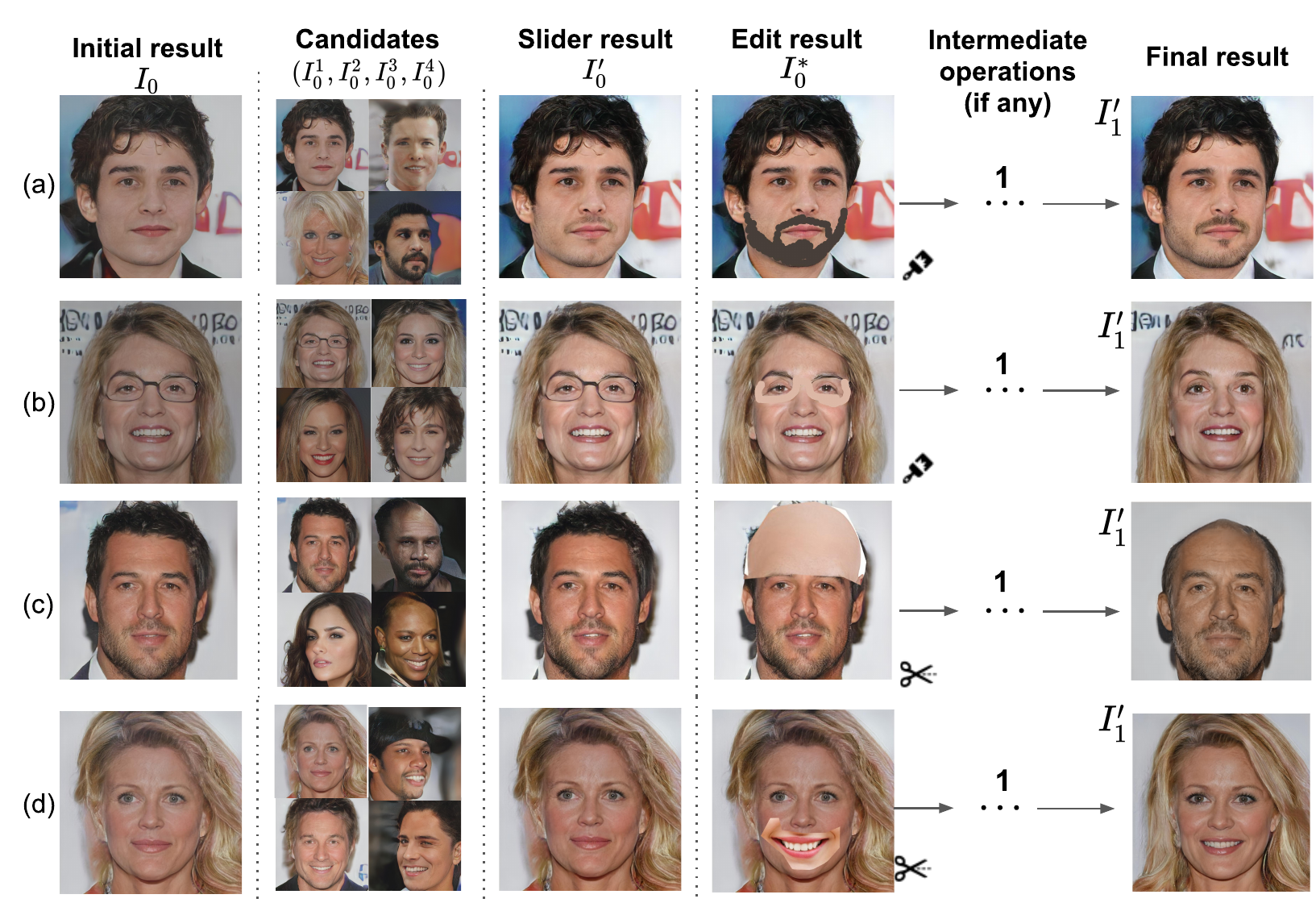}
\caption{
We use our method to control the synthesis results using PGGAN~\protect\cite{Karras2018ICLR}.
The user is able (a) to add beard, (b) remove glasses, (c) make a man bald and old, and (d) make a woman smile.}
\label{fig:example_result_pggan}
\end{figure*}

\begin{figure*}
\centering
\includegraphics[width=0.95\linewidth]{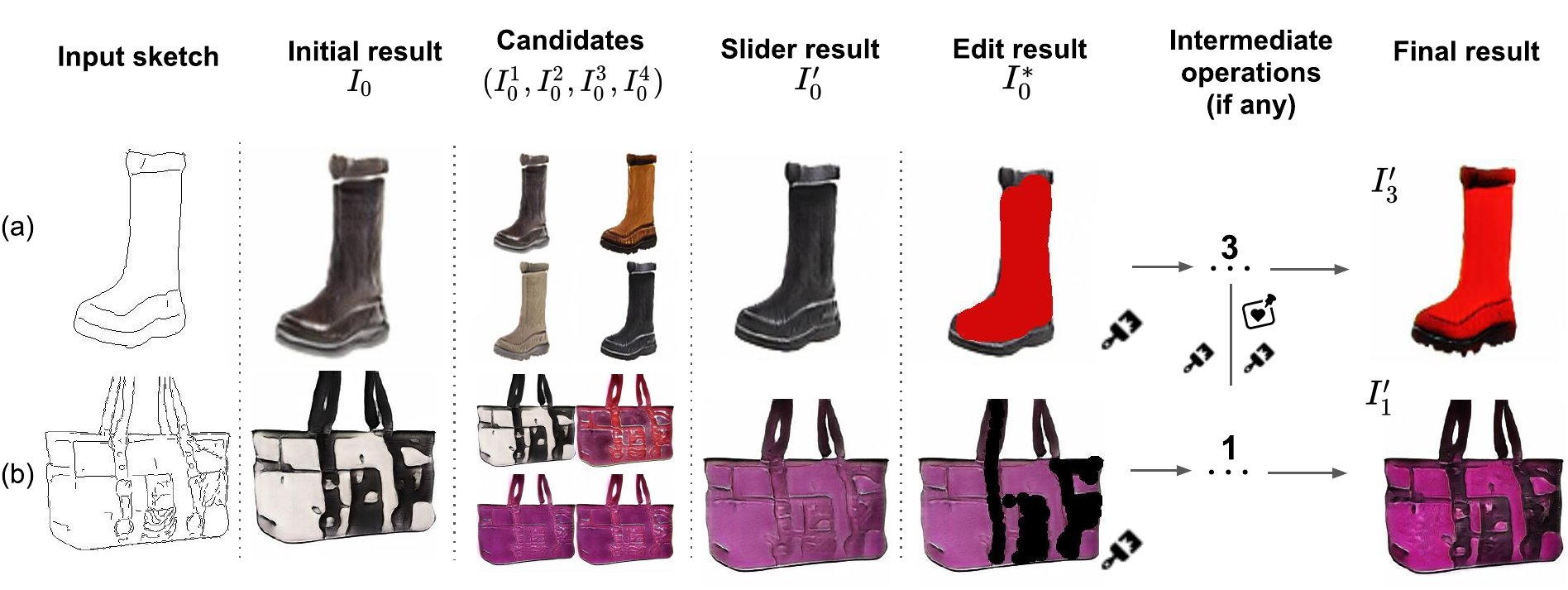}
\caption{
The control sequences of our framework running on MUNIT~\cite{MUNIT}.
The users are able to control the slider and use the color paint tool to design (a) a new boot, and (b) a new handbag. 
}
\label{fig:example_result_munit}
\end{figure*}

\begin{figure*}
\centering
\includegraphics[width=\linewidth]{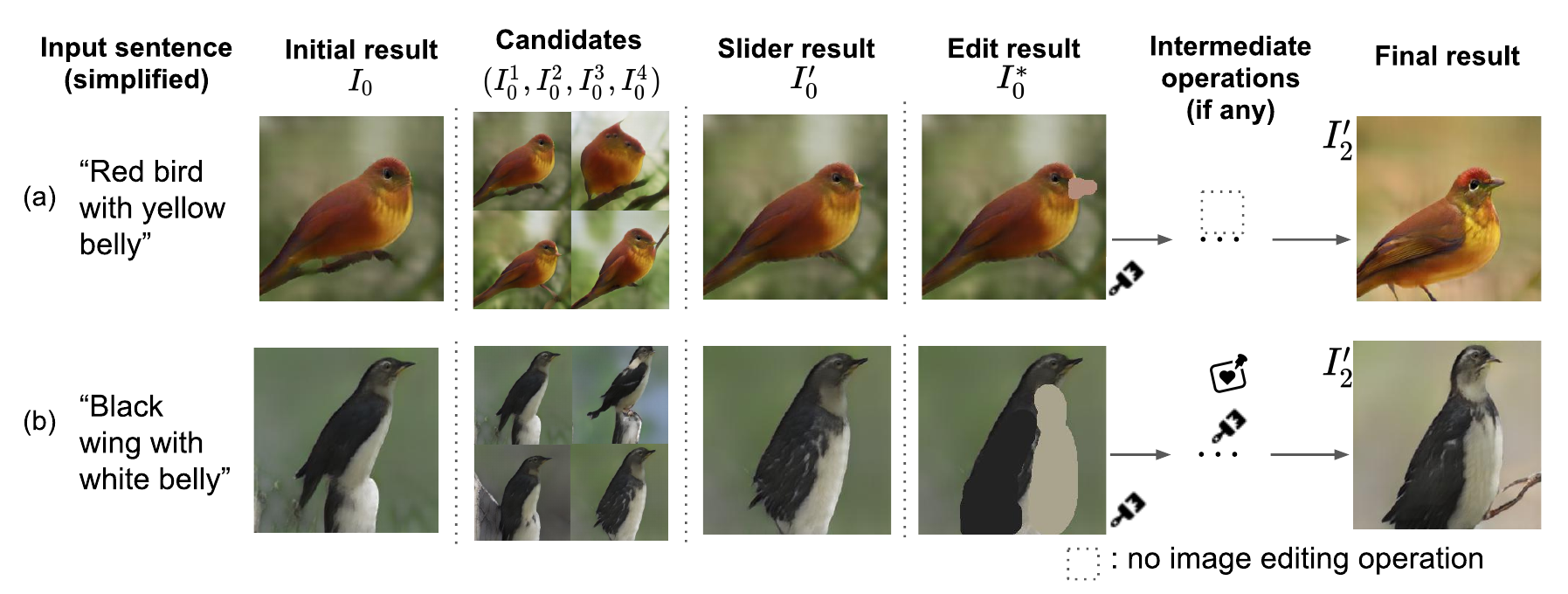}
\caption{
We use our method to control the synthesis results from text-to-image application (using AttnGAN~\cite{attngan}).
The user is able to (a) add a larger beak, and (b) improve the synthesized bird in only few iterations. 
}
\label{fig:example_result_attn}
\end{figure*}

\begin{description}[style=unboxed,leftmargin=0cm]
\item[Image Generation with Random Noise]
First, we applied our framework to PGGAN~\cite{Karras2018ICLR} to control the appearance of human faces.
We used the pretrained generator model provided by the paper author\footnote{https://github.com/tkarras/progressive\char`_growing\char`_of\char`_gans}.
The latent space explored was the random latent vector drawn from the normal distribution $ \mathcal{N}(0,1)$ (the blue latent noise vector in \figname~\ref{fig:network_arch}(a)) fed into the network.
Our method can be used to control different parts of the face; to change hair colors and styles (change hair color and style in \figname~\ref{fig:teaser}(a), and remove hair in \figname~\ref{fig:example_result_pggan}(c)); to add or remove beards (\figname~\ref{fig:example_result_pggan}(a)); to remove accessories such as earrings and glasses (\figname~\ref{fig:example_result_pggan}(b)); and to add or remove smiles (\figname~\ref{fig:example_result_pggan}(d)). 
These examples show that, using simple controls, predictive and high-quality results matching user preferences were possible.

\item[Image Translation]
We applied our method to MUNIT~\cite{MUNIT}, which performs image-to-image translation (\eg~hangbag sketches into photographs of hangbags of different colors).
We used the pretrained models provided by the paper the authors\footnote{https://github.com/NVlabs/MUNIT}.
MUNIT separates content and style into separate latent vectors, and we applied our method to the 8-d style vector (the blue latent vector part of \figname~\ref{fig:network_arch}(b)).
Our method could be used to design new heels (\figname~\ref{fig:teaser}(b)), boots (\figname~\ref{fig:example_result_munit}(a)), and hangbags (\figname~\ref{fig:example_result_munit}(b)) given sketch inputs by assigning different colors for different parts.

\item[Image Synthesis from Text]
Finally we applied our method to AttnGAN~\cite{attngan}, which allows text-to-image synthesis.
We choose AttnGAN as a state-of-the-art text-to-image synthesis method to show that our framework could handle various network architectures.
AttnGAN encodes input text into a sentence vector, accompanied by a random latent vector drawn from a normal distribution $ \mathcal{N}(0,1)$, and synthesizes an image.
We applied our method to the 100-d random noise vector (the blue latent vector part in \figname~\ref{fig:network_arch}(c)).
We showed that our method can be used to improve the synthesized bird images (\figname~\ref{fig:teaser}(c) and \figname~\ref{fig:example_result_attn}(b)), and adding bigger beak (\figname~\ref{fig:example_result_attn}(a)).
\end{description}

\section{Ablation study}
\cgfhl{We conducted two experiments to validate the efficacy of our sequential subspace search with multi-way slider.
}
\subsection{Synthetic function study}
\begin{figure}[h!]
    \centering
    \includegraphics[width=\linewidth]{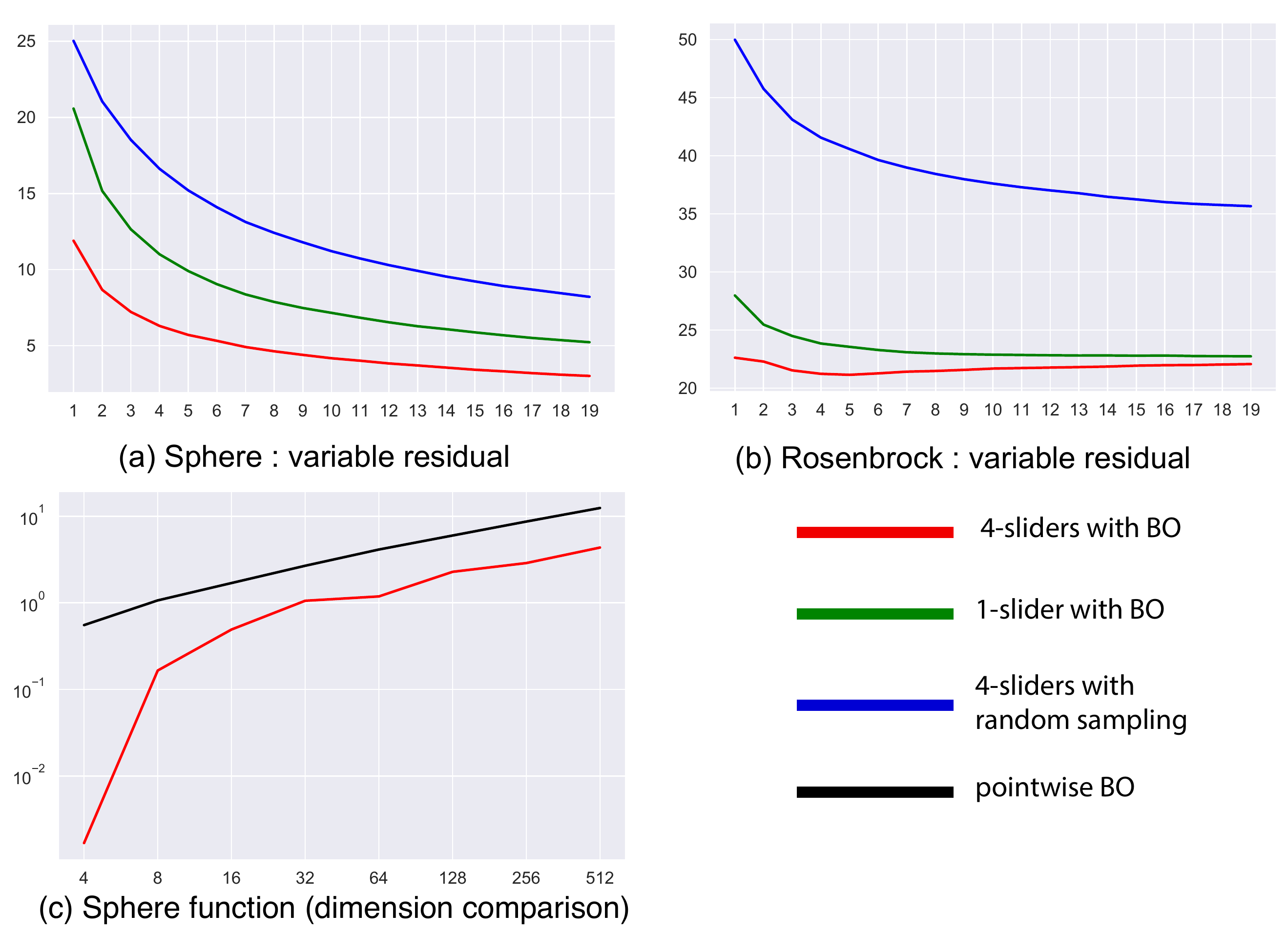}
    \caption{
    We compare the variable residuals (vertical axis) over iterations (horizontal axis) with the two variations of our method.
    The result of Sphere function is shown in (a), and the result of Rosenbrock is shown in (b).
    \cgfhl{(c) We also compare the variable residual per dimension (y-axis) over the number of dimension (x-axis) between 4-sliders BO and point-wise BO. 
    It confirms that higher dimension BO is more difficult to optimize thus require additional user editings to assist the optimization.
    }
    }
    \label{fig:synthetic_fig}
\end{figure}

The first experiment is to validate that the full version of our method can optimize the human preference function more efficiently than the following variations of our method:
\begin{itemize}
\item \textit{random sampling}: at each iteration $i$, we randomly sampled three 512-D latent vectors as $\{z^c_i\}_{c=2}^{4}$ and used a generator to generate the corresponding candidtate faces $\{I^c_i\}_{c=2}^{4}$. 
\item \textit{1-slider}: we used one-slider instead of n-sliders (we use $n=4$ for all of the cases shown in this paper).
\end{itemize}
Following~\cite{Koyama:2017:SLS}, we assumed the ground truth of the human preference function to be represented by the following two synthetic functions: (a) the \textit{sphere function}: \mbox{($f(x) = \sum^d_{i=1} x^2_i,$)}, evaluated over the interval \mbox{$x_i \in [-5.12, 5.12]$},
and (b) the \textit{Rosenbrock Valley} function (${f(x) = \sum^{d-1}_{i=1}[100(x_{i+1} - x^2_i)^2  + (1 - x^2_i)^2]}$), evaluated over the interval $x_i \in [-5, 10]$.
The global minimum of the sphere function is $0$ at $\hat{\mathbf{x}}=\mathbf{0}$ and $0$ in the case of the Rosenbrock Valley function at $\hat{\mathbf{x}}=\mathbf{1}$.
We evaluated both functions with $512$d and ran the experiment for 20 iterations.
We recorded the residuals : $r= \|\mathbf{x}_{20} - \hat{\mathbf{x}} \|_2^2$, where $\mathbf{x}_{20}$ is the optimized value after 20 iterations.
\figname~\ref{fig:synthetic_fig}(a) and (b) show the average residuals of the variables.
These results indicate that our method
(4-sliders with Bayesian optimization) converges faster on both functions tested with the same number of iterations.

\cgfhl{
In addition, we also compared between our 4-slider methods with traditional point-wise Bayesian optimization. 
We performed optimization on the \textit{sphere function} under the same condition in the above ablation study test.
We shows the average residuals at iteration 20 (total residual divided by the number of dimension) of the variables in \mbox{\figname~\ref{fig:synthetic_fig}}(c).
These results indicate that our method
(4-sliders with Bayesian optimization) converges faster than pointwise method. In addition, it also demonstrates that higher dimension is harder to optimize for Bayesian optimization.
Our method compensate the difficulties of high dimensional Bayesian optimization by allowing the users to provide image editings.
}

\subsection{Slider-time complexity study}
\cgfhl{
The second experiment investigates how the number of sliders affects the time required for the user to locate the optimal candidate in the subspace formed by all of the sliders.
A potentially overlooked factor of the comparison in the ablation study is that, in each iteration, our method inherently provides more candidates to the user. 
Although multi-way slider widens user selection and was well recognized in the pilot user study, it increases the user's cognitive load and is likely to spend more time to manipulate the multi-way slider.
We conducted a study to address this factor further to provide a more rigorous justification for using the multi-way slider.
}
\subsubsection{Procedure.}
\cgfhl{
In this study, we want to show that the user can explore the same target latent space (built by the generator of each application) more efficiently with $4$-sliders and the corresponding high-dimensional search space compared to $1$- slider and its' 1D search space.
In total, we recruit 3 participants, with 2 males and 1 female.
Each participant is asked to reach the provided reference image using both $1$-slider and $4$-sliders version of our interface.
We disable all the image editing tools in our interface and randomized the order of using two versions of the interface.
Disabling the image editing tool is to evaluate the efficacy of exploring the latent space slider using the sliders solely.     
Each of the participants has 5 minutes to practice on using the provided interface.
After practice, we will provide 5 reference images in total to each participant.
We measured the average interaction time for each image at each iteration.
}
\subsubsection{Result.}
\cgfhl{
For each iteration, each participant spent 17.2 seconds and 53.4 seconds for $1$-slider and $4$-sliders user interface on average.
Our result shows that the time required to manipulate $4$-sliders interface grows only approximately linearly even though the size of the search space grows super-linearly.
The result indicates that the users are exploring the space more efficiently with $4$-sliders version than $1$-slider version. 
In the interviews we had with the participants, they generally appreciate the increase in the flexibility provided by the $4$-sliders interface.
}


\section{User study}
\cgfhl{
We conducted two user studies to compared our method with \mbox{iGAN~\cite{zhu2016generative}} with respect to steering the image synthesis process.
In both studies, we used face image editing with a reference image as the target task.
The only difference is the source of the reference images.
We used real image (i.e., not generated by generative model) in the first user study, and used synthetic image (synthesized by \mbox{PGGAN~\cite{Karras2018ICLR}}) in the second user study.
In the first user study, our goal is to provide perceptual comparison between our method and iGAN.
And the purpose of using synthesized image in the second user study is to provide quantitative comparison in addition to perceptual comparison.
}

\subsection{User study with real reference}
\subsubsection{Procedure}
We randomly sampled four images from a subset of the \mbox{CelebA dataset~\cite{liu2015faceattributes}} and excluded the images used to train \mbox{PGGAN~\cite{Karras2018ICLR}}.
\cgfhl{
Noted that these images are real images not synthetic images.
We use synthetic images to evaluate our method numerically in \mbox{\secname~\ref{syn-target-study}}.
}
In the setting used to test our method, each participant was shown a random initial image and a reference image, and he/she was asked to convert the initial image into the reference image using the system provided.
In the setting used to test iGAN, each participant was shown a blank canvas and a reference image, and he/she was asked to reproduce the reference image.
Participants were judged to have finished the task when any of the below conditions was fulfilled: (1) they were satisfied with the result, (2) they found it hard to improve the results, or (3) the 10 minute time limit was reached.
We recorded the time taken and the final edited image for a further image comparison study.

\myhl{
In total, we recruited 8 people for the user study.
Of the participants, 3 were female and 5 were male.
We divided the participants randomly into two groups, each containing 4 people.
The participants in each group were shown the same two images (image 1 and image 2), and they were tasked with rendering the reference image using both our method and iGAN.
Our user study proceeded as follows: (tutorial A $\rightarrow $ A1$\rightarrow $ A2$\rightarrow $ tutorial B$\rightarrow $ B1$\rightarrow $ B2), where A1 denotes as using method A on image 1.
We randomly swapped the order in which the participants used iGAN and our method to prevent bias (i.e.~half of the participants used iGAN first and the other half used our method first).
Before the participants started to edit the test image, we provided a 10 minute tutorial session on both methods.
From both methods, We removed an import function (``copy and paste'' in our method and ``import'' in iGAN), which users can use to ``cheat'' by asking the algorithm to essentially perform manifold projection (i.e.,~optimize automatically for the result without using slider manipulation or drawing).
This provides a better simulation of the actual use case, in which users have no access to any picture of the desired goal, but only a vague idea in their minds. 
}

\vspace{-2.0mm}
\subsubsection{Crowdsourced Evaluation}
After this user study, we carried out a crowd-sourced comparison study to evaluate the visual quality of the resulting images.
For each query, we showed crowdworkers a reference image, the edited result using our method, and the edited result using iGAN.
We then asked them to select the result that better matcheed the reference image.
We composed a survey of 16 queries.
We used Amazon Mechanical Turk interface to conduct the survey, in which each participant was shown all 16 queries, with each query shown twice and the order of the candidates switched.
An image of the user interface used by the Amazon Mechanical Turk workers is in the supplemental material.
For each crowdworker, we discarded inconsistent answers, where a duplicated query was answered differently, and discarded all answers from participants who answered over 25\% of queries inconsistently.
In total, we report the results obtained from 50 crowdworkers that passed our consistency checks.

\begin{figure}[htb]
    \begin{minipage}[t]{.48\linewidth}
        \centering
        \includegraphics[width=\textwidth]{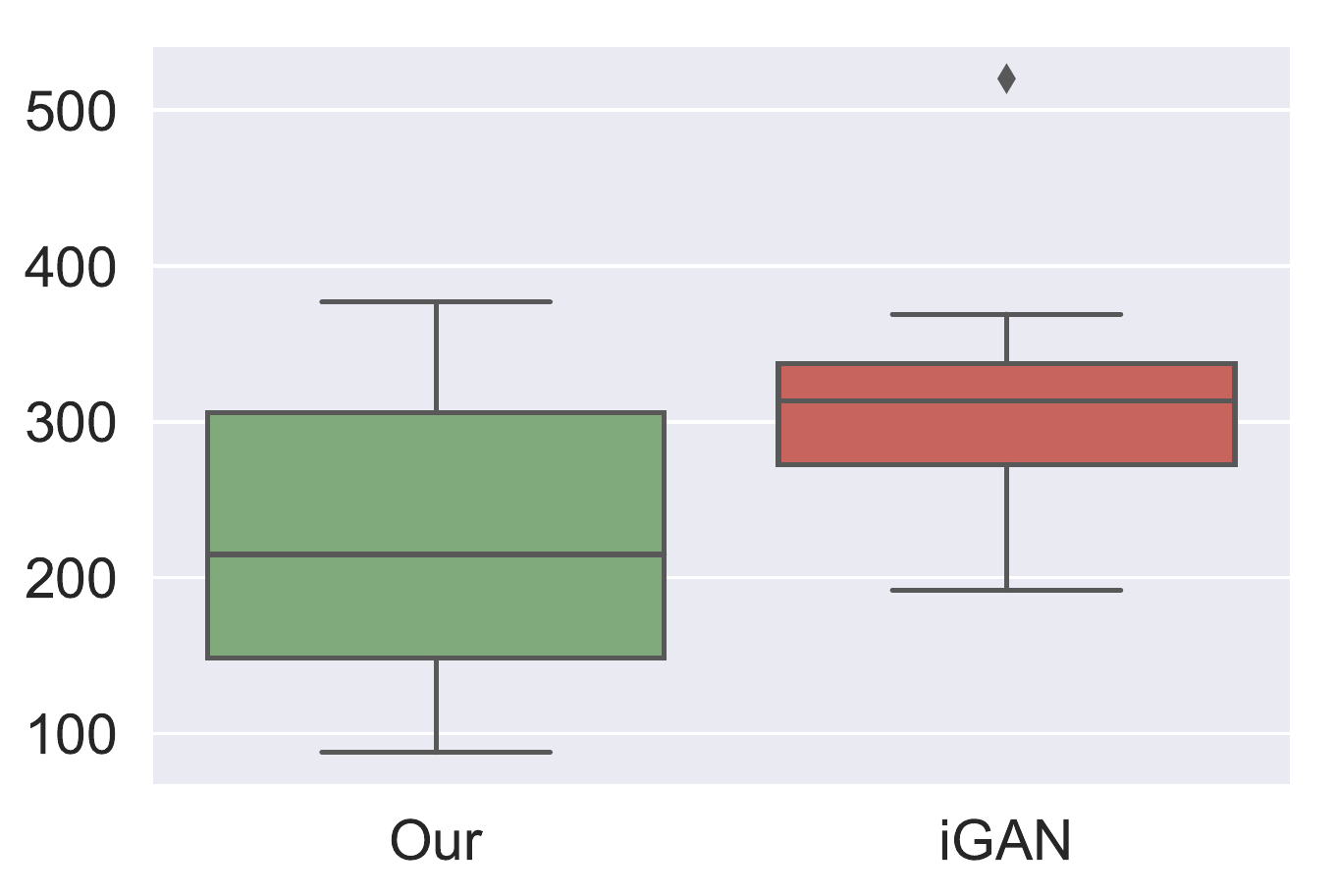}
        \subcaption{
        Average interaction time for all participants, referring to the time the user interact with the system, excluding wait time for computation.}
        \label{fig:study_time}
    \end{minipage}
    \hfill
    \begin{minipage}[t]{.48\linewidth}
        \centering
        \includegraphics[width=\textwidth]{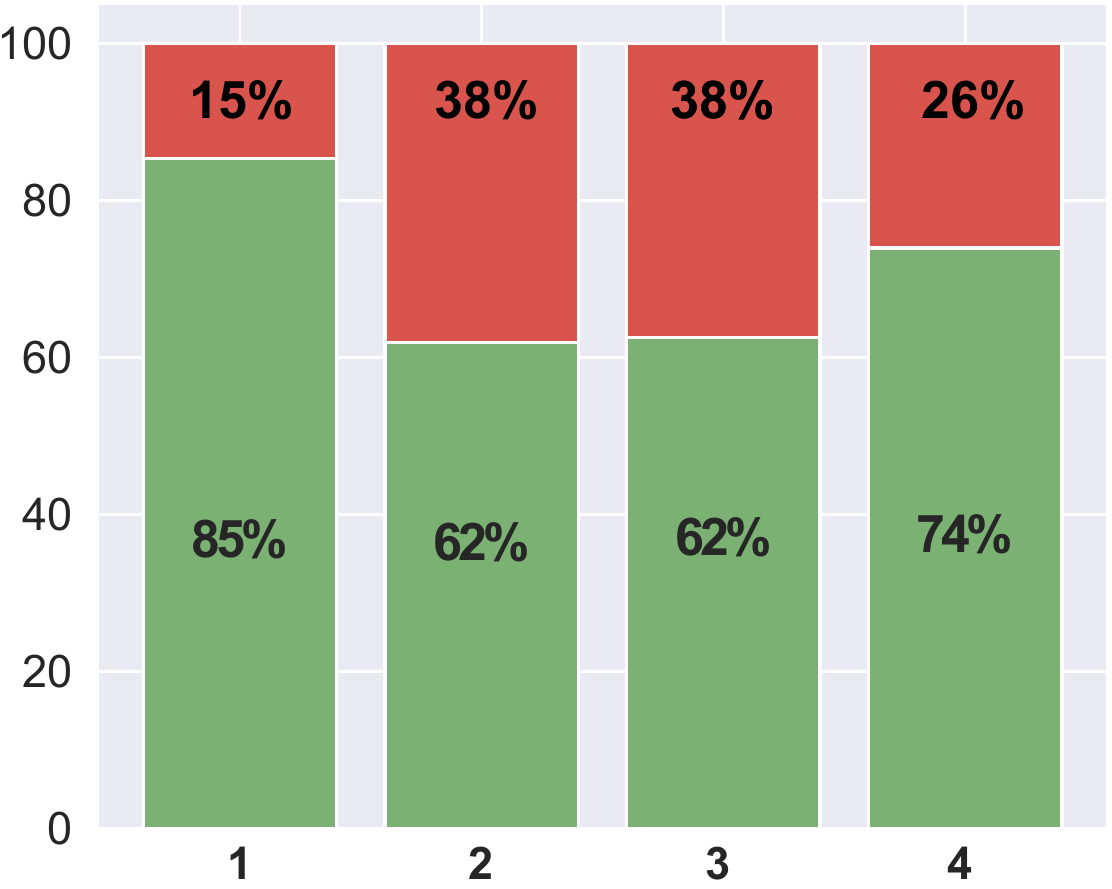}
        \subcaption{
        The percentages of crowdworker preference to our method (green) and iGAN (red) on different reference images (shown in \figname~\ref{fig:study_results}).
        }
        \label{fig:study_error}
    \end{minipage}  
    \label{fig:study_img_diff}
    \caption{
    Some results from our user study. 
    (a) We show the interaction time for both our method and iGAN, and (b) the results of the crowd-sourced comparison user study across four reference images.
}
\label{fig:study}
\end{figure}

\vspace{-2.0mm}
\subsubsection{Study Result}
\begin{figure*}
     \centering
     \includegraphics[width=\linewidth]{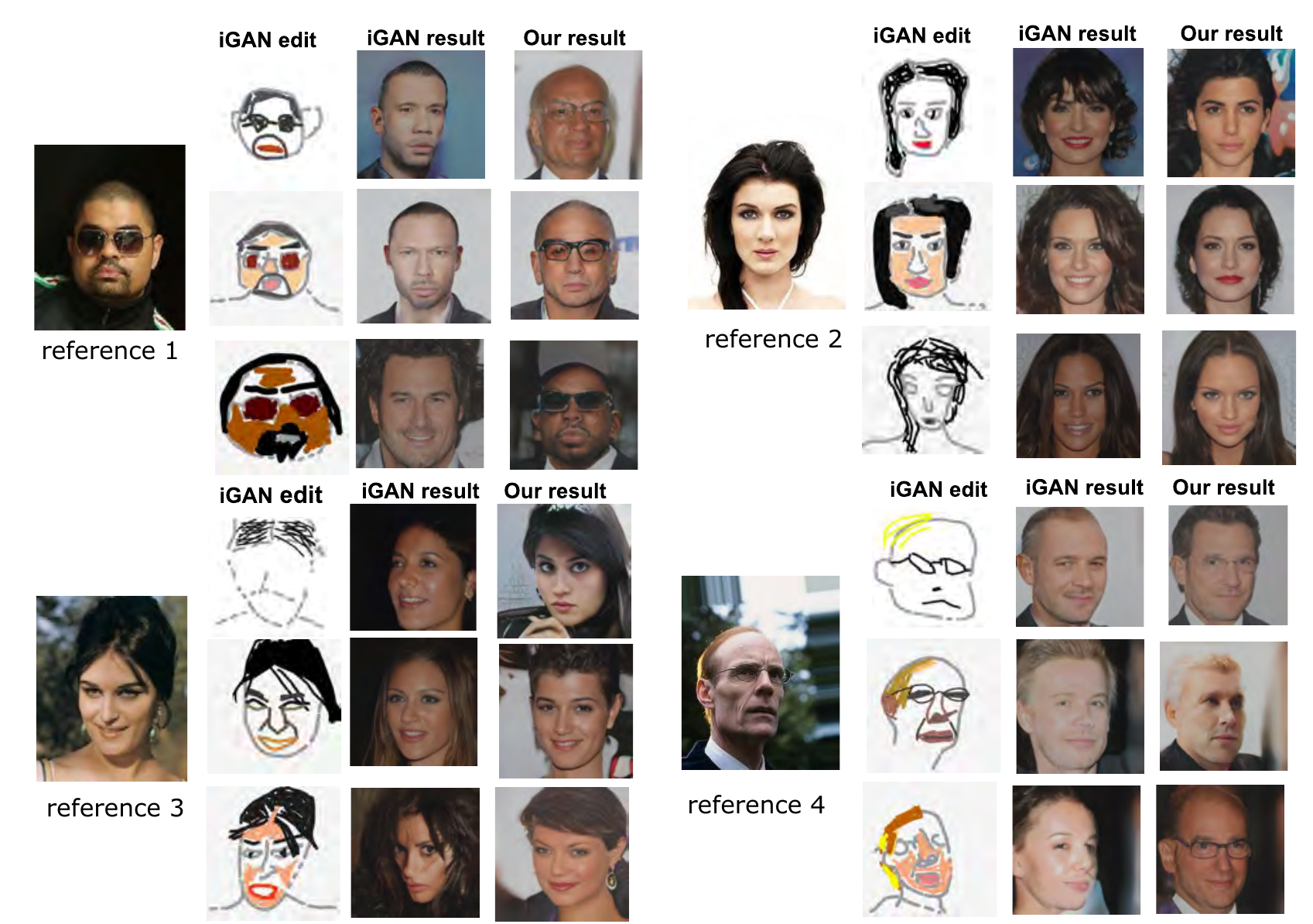}
     \caption{
     Here we show several edited results from user study.
     Grey lines in iGAN edit represent contour sketching and the rest strokes represent colour painting.
     (Please find more in supplemental material).
     }
     \label{fig:study_results}
 \end{figure*}

We compared the user interaction time for both methods, and the results are shown in \mbox{\figname~\ref{fig:study}(a)}.
The user interaction time was significantly shorter using our method than using iGAN.
\mbox{\figname~\ref{fig:study}(b)} shows the results of the crowd-sourced comparison user study.
As shown, the results obtained using our method were voted higher than those obtained using iGAN across all four reference images. 
In total, 77\% of the crowdworkers preferred our results to the results obtained using iGAN (we show the voting statistics for each query in the supplemental material).
The edited images obtained from the user study are shown in \mbox{\figname~\ref{fig:study_results}}.
\cgfhl{
Moreover, the number of sliders the participant used for each iteration is $2.44$ on average (the separate average numbers are $2.25, 2.3, 2.83, 1.9, 2.27, 3.2, 2.5$ and $2.27$).
We provide the slider values for all the participants in a separate spreadsheet file as supplemental material. 
}

\cgfhl{
We attribute the differences between our results and iGAN's results to the fact that novice users often fail to render their mental images accurately using drawing tools. 
This makes direct optimization, such as iGAN, perform poorly in practical tasks. 
Alternatively, our method provides an additional slider exploration mechanism and it reduce the burden for users to explore and express their requirements.
Similar to our method, \mbox{iGAN} provides multiple candidates from which the user can choose.
However, these candidate images tent to converge to a single image after several iterations since they are obtained by optimizing the same loss function.
In our user study, we observed that \mbox{iGAN} often provides very similar candidates that are of little use to the user.
}

\begin{figure*}[h!]
    \centering
    \includegraphics[width=\linewidth]{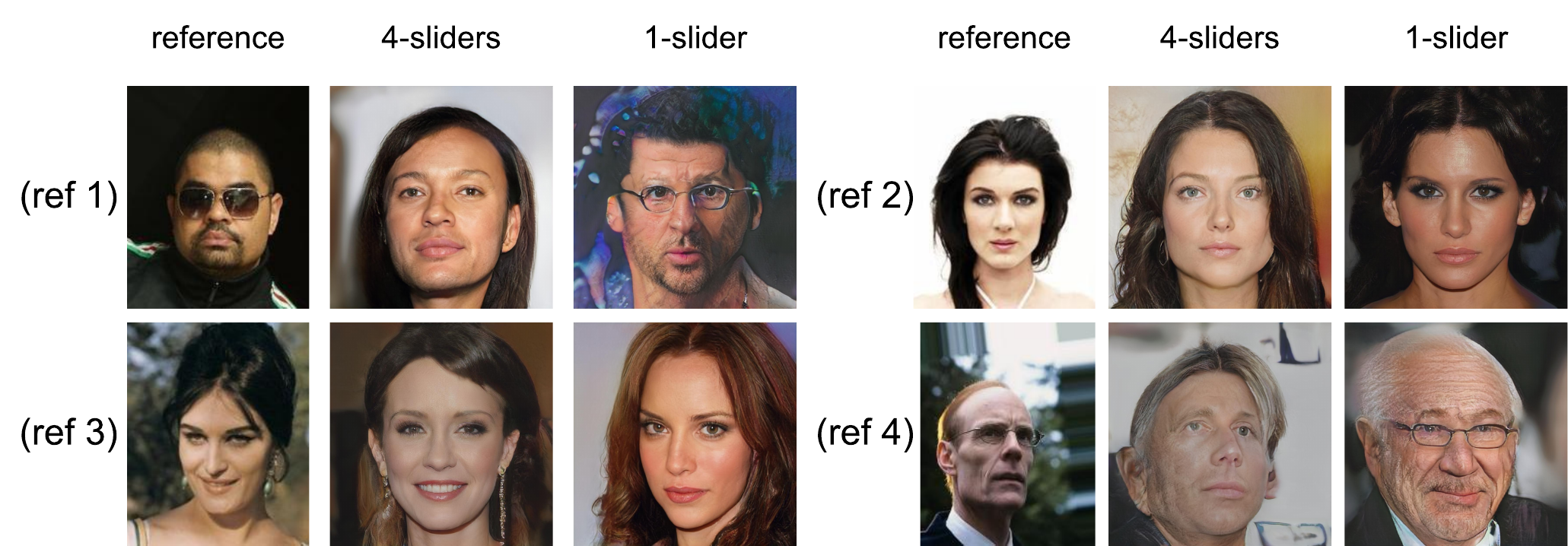}
    \caption{
    We show the results of the additional user study comparing 4-sliders v.s. 1-slider of our method.
    We use the same reference images we used in the user study comparing with iGAN.
    }
    \label{fig:slider_count_test_result}
\end{figure*}

\cgfhl{
Finally, we recruited another 4 participants and conducted an additional study to compare 1-slider v.s. 4-sliders version of our method.
The procedure is the same as the original user study but we only asked each participant to work on one reference image.
The results are shown in \mbox{\figname~\ref{fig:slider_count_test_result}.}
All participant use 15 iterations of 1-slider version of our method, and 4 iterations of 4-sliders version.
The purpose is to participants use the same amount of sliders while they finish the task using two versions of our method. 
}



\subsection{User study with synthetic reference}\label{syn-target-study}
\subsubsection{Procedure}
\cgfhl{
In the previous study we show that our method can create image that is perceptually more similar to the reference image than iGAN.
However, to further evaluate our method quantitatively, 
we conducted a simulated test which the user preferred image ($I_{sim}$) is indeed generated by the generator, \mbox{\ie~$I_{sim} = G(z_{sim})$}.
The procedure of this user study is identical to the original main user study, \mbox{\ie~we asked the user to reach $I_{sim}$ using the system provided}.
After we obtain the user-edited image (${I}' = G({z}')$), we compute the distance $d = \|z'- z_{sim}\|_2$ as quantitative measurement.
The rationale behind this study is that the generators of most of the state-of-the-art methods are still vulnerable from artifacts and can not capture the data distribution perfectly, this task simulates the situation when we have perfect generator that can generate any image.
Given this observation, we choose to use generated image ($I_{sim}$) as the reference image for meaningful quantitative measurements.
}

\subsubsection{Study Result}
\begin{figure}[ht!]
    \centering
    \includegraphics[width=\linewidth]{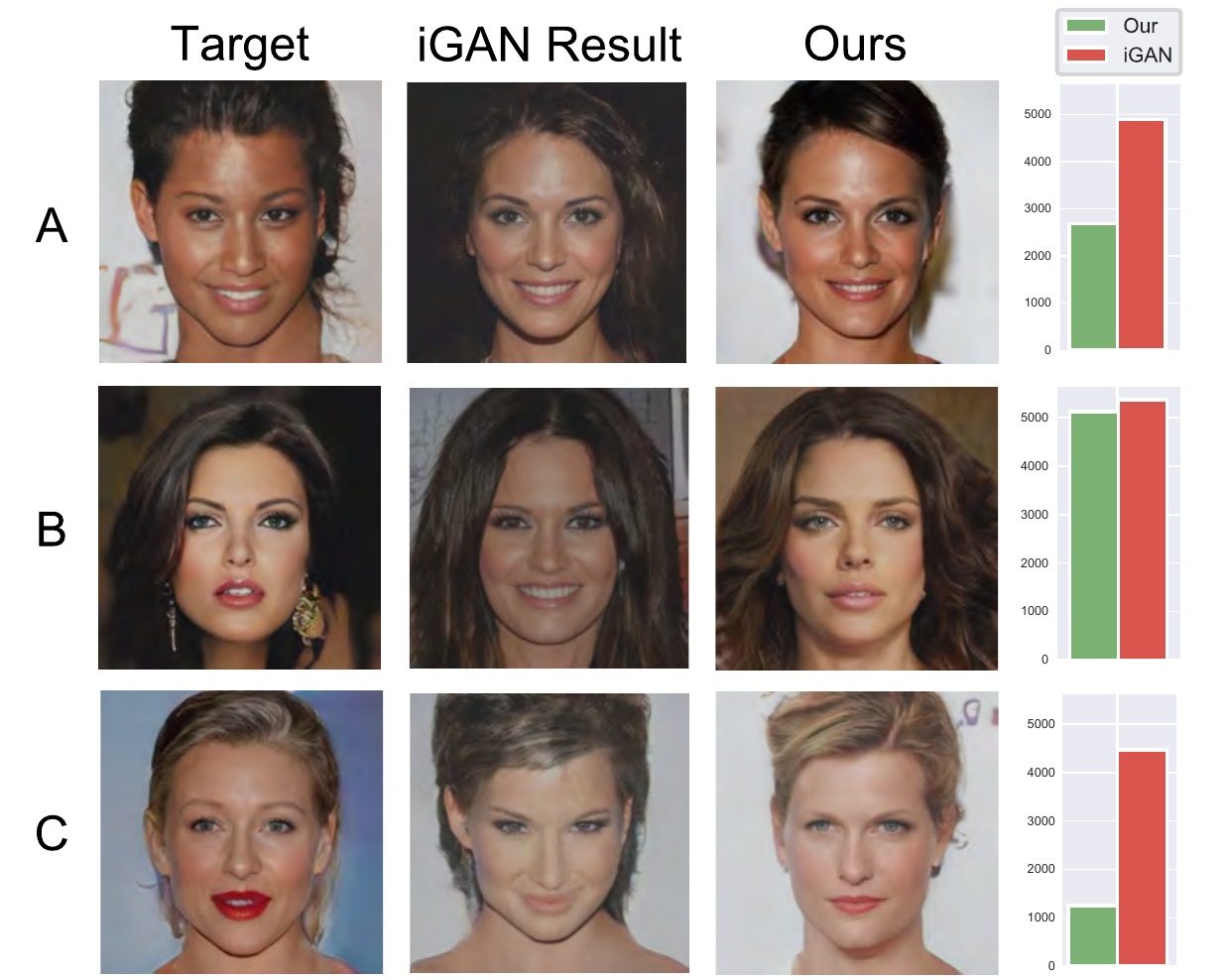}
    \caption{
    On the leftmost column we show the randomly chosen reference images shown to the participants.
    And we show the results generated using iGAN (2nd column) and the result generated using our method (3rd column).
    And we show the euclidean distances between the images generated by the our system and iGAN measured in latent space in the last column.
    Lower distance implies the generated result is closer to the target in latent space.
    For case A, our result generally matches the head pose, skin tone, hair style and mouth shape compare to iGAN result. 
    For case B, our result matches the target with hair style, mouth shape and skull structure, yet iGAN result is closer in terms of head pose. 
    For the last case, our result is closer in terms of lips color, head pose and hair color.
    }
    \label{fig:yh_study}
\end{figure}

\cgfhl{
In this study we recruited 3 participants.
For each participant, we randomly generated $I_{sim}$ at the beginning of the study.
We show the reference image and the user-edited image using both iGAN and our method in \mbox{\figname~\ref{fig:yh_study}} (2nd and 3rd column). And we measure the distance between the latent vectors as our metric as shown in the last column of \mbox{\figname~\ref{fig:yh_study}}.
The distances $d$ of the $512$-d latent vector for A, B, and C cases using our method are 2681.9409, 5129.3931, and 1253.9229, and the distances using iGAN are 4892.0669, 5366.6924, 4467.7153 respectively.
While it is impossible to draw a definitive conclusion from this study alone, it supports our hypothesis that our method and user interface can indeed steer the user to a certain point-of-interest on the manifold constructed by the generator. 
}




 \begin{figure*}
     \centering
     \includegraphics[width=\linewidth]{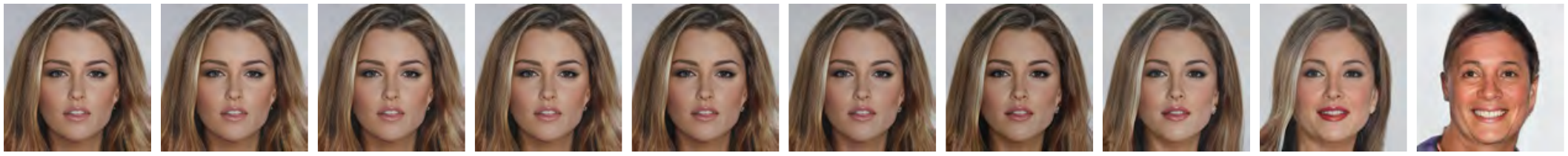}
     \caption{
     The latent space constructed by the generative model contains uneven distribution. 
     Here we samples the linear interpolation between two latent vectors (the left most and the right most). 
     And we can observe that this uneven distribution hinders the ability for user to fully explore the latent space.
     }
     \label{fig:bad_linear_interp}
 \end{figure*}
\begin{figure}
     \centering
     \includegraphics[width=\linewidth]{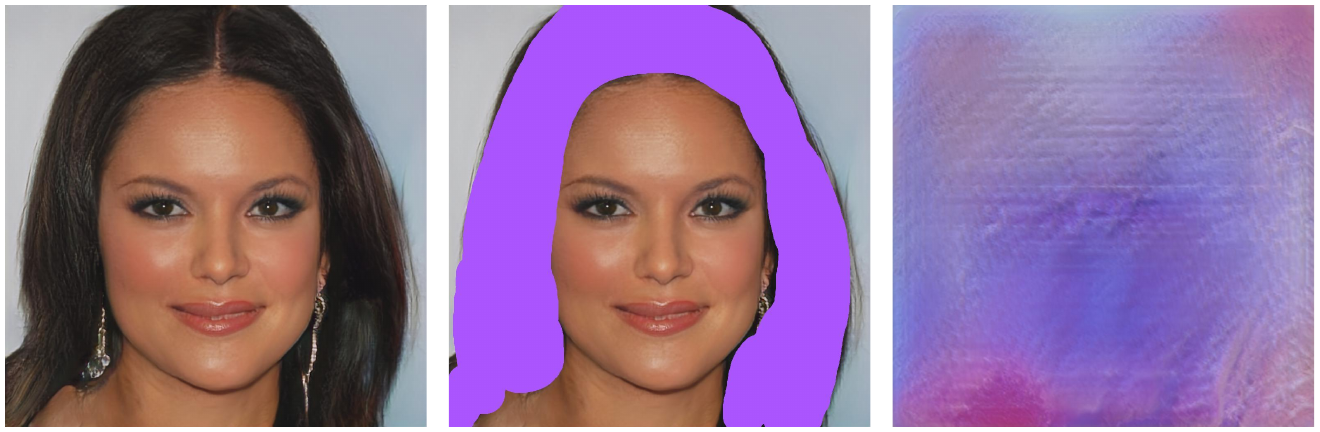}
     \caption{
     If the user input that aim to generate image not within capability of the generator (\eg~the purple hair), the system will fail to optimize for an image that neither match the user's expectation nor resemble a human face.
     }
     \label{fig:bad_input}
 \end{figure}
\section{Discussion, limitations and future work}
Bayesian Optimization (BO) has been proven to be effective in handling moderate-to-high dimensional problem. 
Yet for high dimensional problem, Bayesian Optimization in some cases shows moderate result compared with random search~\cite{highdim-dropout}. 
In our ablation study we showed that our method outperforms random search, even though our method does not have direct observation of the objective function. 
In this section we attempt to provide insights to explain why that is the case from the following two perspectives:

\begin{description}[style=unboxed,leftmargin=0cm]
\item[High implicit observation number]
In our method, the user (or in our ablation study, a local optimizer that explores the entire linear subspace formed by the slider) observes significantly more samples (by manipulating the slider) compared to the amount of sample provided to the optimization algorithm. 
While not directly used during optimization process, we argue that the samples provided to the user contain more information than the ones using traditional value-based Bayesian Optimization.

\item[Dimensional Reduction]
Our method did not perform dimensional reduction explicitly. 
However, we argue that the slider mechanism provides a mapping to reduce the dimension of the optimization problem. 
REMBO~\cite{rembo-wang2016bayesian} also provides a dimensional reduction method (via random mapping) and demonstrates superior results on extremely high dimensional problem (1 billion). 
Their method works on problem with low effective dimensions, \ie~minimal number of dimensions required to represent a space. 
At the same time, we observed that by changing some dimensions of the latent vector often leads to no visible changes on images generated by PGGAN.
This observation suggests that the effective dimension of such generative image modeling model is potentially low, which aligns with the conclusion from REMBO~\cite{rembo-wang2016bayesian} that Bayesian Optimization is able to perform well on such cases where the effective dimension is way lower than the original dimension.
\end{description}

Following are some limitations and future works.
\begin{description}[style=unboxed,leftmargin=0cm]
\item[Bounded by the generator.]
We used an existing generator which limited the quality of the synthesis images.
We used state-of-the-art image generative modeling architecture to perform tasks within the confines of the network; users achieved satisfactory results within a few trials. 
However, if tasks exceeds the bounds of the application-tailored network architecture or the training data used, our framework will fail to optimize user preferences. 
\cgfhl{
The common limitations are: (i) the latent space contains obvious discontinuities so that the subspace for user to explore is degenerated (\mbox{\figname~\ref{fig:bad_linear_interp}}) and (ii) if the user input exceed the capacity of the generator, the optimized result will be broken (\mbox{\figname~\ref{fig:bad_input}}).
}
In future, we are interested in developing indications as to whether
a task is within the boundaries of a network or not.
\item[Improving the generator via interaction.]
We froze the generative model during user interaction, but in future it might be useful to incorporate user interaction to improve the model.
The concept of human-assisted generative model learning was proposed in~\cite{improveGAN}, and the results were meaningful, but the interaction method was not specified.
However, by combining our interactive system with existing learning system, we believe that it may be possible to improve the network via crowd-sourcing or other means. 
Thus, a smaller dataset can be used to train the network and improve synthesized results via human interaction, which is useful when the cost of acquiring a large dataset is high (\eg, medical images). 
\end{description}


\section{Conclusion}
We introduced a human-in-the-loop optimization method that helps users to steer the generative image modeling methods.
Our method provides a multi-way slider interface to further ease the steering process in comparison to a free-form stroke-based drawing interface.
The slider interface prevents users from providing out-of-distribution input.
Our method only requires a pretrained generative model, so that it provides a plug-and-play option for various generative image modeling methods.
It is not limited to the applications described in this paper.
We believe that the proposed method will greatly increase the convenience of generative models, and will lead to novel image generation applications.

\section{Acknowledgement}
This work was supported by JST CREST JPMJCR17A1.
We would like to thank Bing-Yu Chen and anonymous reviewers for insightful suggestions and discussions.
During this work, I-Chao Shen was also supported by the MediaTek Fellowship.

\bibliographystyle{eg-alpha-doi}
\bibliography{ganui_cgf}

\end{document}